\documentclass[12pt]{article}
\usepackage{amssymb}
\usepackage{epsfig} 
\usepackage{epsf} 
\usepackage{graphics} 
\usepackage{graphicx}

\begin{document}

\newcommand{\bc}{\begin{center}} 
\newcommand{\ec}{\end{center}} 
\newcommand{\ikl}{\int_k^\Lambda} 
\newcommand{\dd}{\frac{d^2k}{(2 \pi)^2}} 
\newcommand{\dt}{\frac{d^3k}{(2 \pi)^3}} 
\newcommand{\dtp}{\frac{d^3p}{(2 \pi)^3}} 
\newcommand{\dq}{\frac{d^4k}{(2 \pi)^4}} 
\newcommand{\dn}{\frac{d^nk}{(2 \pi)^n}} 
\newcommand{\PLB}{{\it{Phys. Lett. {\bf{B}}}}} 
\newcommand{\NPB}{{\it{Nucl. Phys. {\bf{B}}}}} 
\newcommand{\PRD}{{\it{Phys. Rev. {\bf{D}}}}} 
\newcommand{\AOP}{{\it{Ann. Phys. }}} 
\newcommand{\MPL}{{\it{Mod. Phys. Lett. }}} 
\newcommand{\del}{\partial} 
 
\begin{titlepage} 
\title {Comparing Implicit, Differential, Dimensional 
and BPHZ Renormalisation} 
\author{{\bf M. Sampaio$^a$},{\bf A.P. Ba\^{e}ta Scarpelli$^b$}, {\bf B. Hiller$^a$}, \\
{\bf A. Brizola$^c$}, {\bf M.C. Nemes$^{c,d}$}, {\bf S. Gobira$^c$} 
\thanks{{\tt E-mails:msampaio@fisica.ufmg.br, scarp@gft.ucp.br, brigitte@teor.fis.uc.pt,
brizola@fisica.ufmg.br, carolina@fisica.ufmg.br, sgobira@fisica.ufmg.br}}\vspace{2mm} \\
{\small {\bf $^a$} University of Coimbra - Centre for Theoretical Physics}\\
 {\small 3004-516 Coimbra, Portugal}\\
{\small{\bf $^b$} Universidade Cat\'olica de Petr\'opolis - 
Rua Bar\~ao do Amazonas, 124 }\\
{\small 25685-070, Petr\'opolis, Rio de Janeiro - Brazil}\\
{\small {\bf $^c$} Federal University of Minas Gerais -
Physics Department - ICEx }\\
{\small P.O. BOX 702, 30.161-970, Belo Horizonte MG - Brazil}\\
{\small {\bf $^d$} University of S\~ao Paulo - Physics Department}\\
{\small P.O.Box 66318, 05315-970, S\~ao Paulo - SP, Brazil}}  
\maketitle

\begin{abstract} 
\noindent 
We compare a momentum space  implicit regularisation (IR) framework with 
other renormalisation methods
which may be applied to dimension specific theories, namely
Differential Renormalisation (DfR) and the BPHZ formalism. In particular, we define 
what is meant by minimal subtraction in IR in connection with DfR and dimensional
renormalisation (DR) . We illustrate with the calculation of the gluon
self energy a procedure by  which  a
constrained version of IR automatically ensures gauge
invariance at one loop level and handles infrared divergences in a
straightforward fashion. Moreover, using the $\varphi^4_4$ theory setting 
sun diagram  as an example and comparing explicitly with the BPHZ framework, we 
show that IR directly
displays the finite part of the amplitudes. We then construct a
parametrization for the ambiguity in separating the infinite and
finite parts whose parameter serves as renormalisation group
scale  for the Callan-Symanzik equation. Finally we argue that
constrained IR, constrained DfR and dimensional reduction are 
equivalent within one loop order.

\end{abstract} 
Pacs:11.10.Gh, 11.25.Db, 11.15.Bt. \\
Keywords: Regularisation methods, Dimension Specific 
\end{titlepage}

\section{Introduction}

It is well known that the higher the symmetry degree of a quantum
field theoretic model the more stringent are the constraints on a
consistent regularisation scheme to handle the divergences which 
appear in diagrammatic expansions. For example, whereas a sharp cutoff
may be successfully employed in most scalar field theories to reflect
the correct physics in perturbation theory, it does not work so well
already for abelian gauge field theories. For gauge field theories, 
DR is one of the most suitable frameworks because the amplitudes 
can be renormalised using for instance a minimal subtraction scheme 
(MS) and readily satisfy the Slavnov-Taylor identities.

However some  symmetries which are present in the integer dimension
may not have a direct analogue in $n$ dimensions. This is the case of
supersymmetric (SUSY), chiral and topological field theories (the so
called dimension specific theories). Some modifications of DR can be
effected in order to mend  certain shortcomings. For instance one may
construct an extension of the algebra of the $\gamma^5$ matrix to
dimension $n$ and control eventual spurious anomalies by imposing the
Slavnov-Taylor identities as constraint equations. This is the usual
procedure in the Electroweak Sector of the Standard Model. In
Chern-Simons theories it may be necessary to employ an hybrid
regularisation procedure by adding Higher Covariant Derivative terms
in the Lagrangian which improves the ultraviolet behaviour. The
remaining divergences are dealt with DR and adopting an extension of
the Levi-Civitta tensor algebra to be compatible with analytical
continuation on the space-time dimension. The main drawback in the
example above is that the calculation may become extremely complicated
beyond the one loop order. A variant of DR called Dimensional
Reduction (DRed) was proposed by Siegel \cite{SIEGEL1}. The latter
differs from DR in the sense that the continuation from $4$ to $n$
dimensions is made by compactification. Thus whereas the momentum (or
space-time) integrals are $n$-dimensional, the number of field
components remains unchanged. Such procedure, however, may introduce
ambiguities in the {\it{finite}} parts of the amplitudes as well as in
the divergent parts in high order corrections. DRed has been largely
employed especially in supersymmetric models as the invariance of the
action with respect to SUSY transformations holds in general only for
specific values of $n$. Unfortunately DRed appears to work well only
at one loop level. In fact, DRed can be shown to be inconsistent in
general with analytical continuation \cite{JACK}, \cite{SIEGEL2} when
$\gamma_5$ matrices and $\epsilon_{\mu_1 \mu_2 \ldots}$ tensors are
considered. In general a pragmatic attitude is
adopted in handling the shortcomings brought by flawed regularisation
frameworks especially when the model in consideration is known to be
free of anomalies. In other words the task of treating the infinities
in diagrammatic calculations, especially for theories which are
sensitive to dimensional continuation, without introducing ambiguities
steming from the regulator employed (that is to say, a regulator
independent method) is still a subject of major interest. Ultimately
it is desirable to construct a framework in which one has
simultaneously : 1) no need to add structure to the Lagrangian and
hence complicate the Feynman rules;  2) (nonabelian) gauge invariance
is systematically guaranteed 
without having to be  imposed as constraint equations order by order;
3) control upon infrared divergences without introducing additional 
machinery and 4) a
method that is friendly from the calculational viewpoint.

The task of treating the ultraviolet infinities in a regulator
free fashion has been firstly  conceived within the BPHZ formalism 
\cite{BPHZ}. This framework relies ultimately in Weinberg's theorem 
which states that a Feynman graph converges if the degree of 
divergence of the graph and all its subgraphs is negative.  
A systematic implementation of this idea is the Dyson's scheme 
which is based on the idea that differentiation with respect to 
the external momentum turns the graph less divergent. Hence in 
Dyson's method the divergent parts of a graph $G$ are subtracted 
by applying Taylor operators $t^{d(\gamma)}$ where $d(\gamma)$ is 
the degree of superficial divergence starting from the smallest 
subgraphs. When overlapping divergences occur care must be exercised
 in such subtraction procedure. The BPHZ framework is the 
generalisation of the Dyson procedure to include overlapping 
diagrams by means of a well prescribed formula called the forest formula. 
Although BPHZ is a very powerful framework which enables 
to construct proofs of renormalizability to  all orders, gauge invariance and hence the 
Slavnov Taylor identities should be imposed as constraint  equations. The reason why 
gauge invariance is broken  when the BPHZ method is applied to nonabelian gauge 
theories lies in the subtraction process which is  based on expanding around an 
external momentum and  thus modifying the structure of the corresponding 
 amplitude. Some modifications in the BPHZ framework  (Soft BPHZ Scheme) must be 
made to handle infrared  divergencies because in the original formulation the  subtraction 
is constructed at zero external momentum  \cite{SOFT}. 

DfR \cite{FREEDMAN}-\cite{PV4} and IR (please see \cite{OAC}-\cite{PRDOC}
 for applications) seem to be very promising in this sense since they do 
not modify the space time dimension or introduce an explicit regulator at
 any step of the calculation. The former is  position space method (contact
 with momentum space is made by means of Fourier transforms) whereas the 
latter is essentially constructed in momentum space. We shall discuss 
these methods in greater detail throughout this paper.  We believe that
 the comparison which we shall outline here will show that IR is a promising
 candidate for handling divergences in field theoretical calculations 
(UV and infrared) in general in a symmetry preserving fashion yet being 
simple from the computational point of view.
      
This paper is organized as follows:  in section 1) we give a brief 
description of DfR and IR and compare  with DR. We work out a few 
examples in $\phi^4$ theory and $QED$  
where we discuss the role played by momentum routing invariance 
in connection with gauge invariance to effect a constrained version
 of IR. We also claim that to one loop order dimensional Reduction, 
DfR and IR are equivalent and define what is meant by minimal subtraction in IR.   
In section 2) we compute explicitly the setting sun diagram in both 
BPHZ, IR and compare with DfR. This is a nontrivial example because it 
possesses an interesting divergent structure from which we will clearly 
see the advantages of applying IR and DfR especially in obtaining the 
finite part. In section 3) we calculate the gluon self energy in $QCD$ within  
IR to show that it can consistently handle the infrared divergences as
 well as readily display the finite part expressed by a class of well 
defined functions. We conclude by outlining a few applications in which 
IR could be useful and perhaps more advantageous.  
   
\section{Relationship between Differential, Implicit and Dimensional Renormalisation}   
   
DfR was introduced by Freedman, Johnson and Latorre \cite{FREEDMAN} as
 a method of regularization and renormalization  in coordinate (Euclidian)
 space. The idea is that the product of propagators is not a distribution 
and so it has no Fourier transform. In DfR,  renormalization is the procedure 
which extend products of distributions into distributions by substituting
 bad-behaved expressions by derivatives of well-behaved ones \cite{HAAGENSEN}
 which are understood as distributions, that is to say, the derivatives are 
meant to act on test functions. It automatically delivers finite Green's 
functions  (which are identical to the bare ones for separate points but 
behave well enough at coincident points) without introducing an intermediate
 regulator or counterterm.  For instance, suppose that we have<B></B> an   
amplitude proportional to the product of massless propagators   
\begin{equation}   
\Delta_0(x) = \frac{1}{4 \pi^2 x^2} \,\,\, , x^2 = x_\mu x_\mu .   
\label{eq:mlprop}   
\end{equation}   
Although $(\ref{eq:mlprop})$ is a well defined distribution  its square 
is not.  According to DfR we search for $G$ such that   
\begin{equation}  
\frac{1}{x^4} = \square G(x^2)   
\label{eq:de}   
\end{equation}    
which also guarantees manifest Euclidian invariance. In solving 
such differential equation we gain  arbitrary scales among which 
 $M$, which is introduced for dimensional reasons in   
\begin{equation}  
G(x^2) = -\frac{1}{4}\frac{\ln x^2 M^2}{x^2} \, ,   
\label{eq:g}   
\end{equation}   
can be shown to play the role of scale variable in the (Callan-Symanzik)
 renormalisation group equation satisfied by the renormalised amplitude. 
The latter is constructed by substituting the l.h.s. of $(\ref{eq:de})$ with 
its r.h.s. where $G$ given by $(\ref{eq:g})$ , that is   
\begin{equation}  
\Delta_0^2(x) \rightarrow \Bigg[ \Bigg( \frac{1}{4 \pi^2 x^2}\Bigg)^2\Bigg]_R 
= -\frac{1}{(4 \pi^2)^2}\frac{1}{4}\square \frac{\ln x^2 M^2}{x^2} \, .   
\label{eq:dfren}   
\end{equation}   
Now $G(x^2)$ does have a Fourier transform, namely $\pi^2 \ln \Big( p^2/\bar{M}^2\Big)/p^2$   
which enables us to write the Fourier transform of $(\ref{eq:dfren})$ as (in the Minkowski space)   
\begin{equation}  
\Bigg[\int_k \frac{1}{k^2 (k-p)^2}\Bigg]_R =  b \ln \frac{\bar{M}^2}{p^2}   
\label{eq:ft}   
\end{equation}   
where    
\begin{equation} 
b = \frac{i}{(4 \pi)^2} \,\, ,  
\end{equation}    
$\int_k \equiv \int d^4k/(2 \pi)^4$ and $\bar{M} \equiv 2 M/\gamma_E$,
 $\gamma_E$ being the Euler's constant. A comparison between DfR and 
DR's can be easily made. For the sake of clarity, we briefly outline 
it here for the case of massless theories following \cite{DUNNE}, \cite{PV3C}.    
   
Power law singularities of the type $|x|^{-n}$ cannot have
 their degree of divergence decreased by using the identity   
\begin{equation}  
|x|^{-p} = \frac{\square |x|^{-p+2}}{(-p+2)(n-p)}   
\label{eq:id}   
\end{equation}   
and setting $p=n$ because of the pole $1/(n-p)$. Alternatively we 
may try and regulate by dimensional continuation moving away from $n$ by    
$-r \epsilon$ and thus using $(\ref{eq:id})$ to get   
\begin{eqnarray}   
\mu^{r\epsilon}|x|^{-n+r\epsilon} &=& {1\over   
\epsilon} \mu^{r\epsilon} {1\over r(2-n+r\epsilon)}\square |x|^{-n+r\epsilon +   
2} \nonumber \\   
&=& -{1\over \epsilon}  {4\pi^{n/2}\over r\left( 2 - n+r\epsilon\right) \Gamma   
\left( {{n\over 2}}-1\right)} \delta^{(n)} (x) \nonumber \\
&+& {1\over   
2(2-n)}\square \Big( {\ln \mu^2 |x|^2\over |x|^{n-2}}\Big) + {\cal   
O}(\epsilon)\,\, .   
\label{eq:rd}   
\end{eqnarray}   
Thus in the dimensional approach the singularity $x=0$ is regulated 
by an infinite counterterm proportional to $\delta^{(n)}(x)/\epsilon $.
 There is also a ${\cal{O}}(\epsilon^0)$ piece in the term proportional 
to  $\delta^{(n)}(x)$ (finite counterterm). If we subtract such 
counterterms we will be left  with a term which is just the result 
obtained within DfR after identifying $M$ with $\mu$. Alternatively 
we can use ($\ref{eq:rd}$) to compute the regularised value of 
$\Delta_0^2(x)$. Given that the massless propagator in $n$ dimensions
 is $\Delta_0(x) = -\Gamma (n/2-1)|x|^{2-n}/(4 \pi^{n/2})$ and 
$\square \Delta_0(x) = \delta ^{(n)}(x)$ we  have   
\begin{eqnarray}  
& &\mu^{2 \epsilon}\frac{\Gamma^2(n/2-1)}{4^2 \pi^n}x^{4-2n}= 
\frac{1}{(4\pi^2)^2}\Bigg[ \pi^2 \frac{1}{\epsilon} 
\delta^{(n)}(x) \nonumber \\ &-& \frac{1}{4} \square 
\frac{\ln (x^2 \mu^2 \pi \gamma_E e^2)}{x^2}\Bigg] + {\cal{O}(\epsilon )} \, .   
\label{eq:rd1}   
\end{eqnarray}   
Some comments are in order: The set of rules of DfR \cite{PV1} which 
fix local counterterms to establish equation $(\ref{eq:dfren})$ is 
called constrained differential renormalization (CDfR). In particular, 
in CDfR one does not introduce arbitrary constants for singular behaviour
 worse than $x^{-4}$. CDfR can be shown to implement gauge invariance  
automatically at least to one loop order. From  $(\ref{eq:rd1})$ it is clear that CDfR and   
DR's with a fixed initial condition given by  $(\ref{eq:rd1})$  
are equivalent under a MS scheme upon the identification $M^2 = \mu^2 \pi \gamma_E e^2$.
  As a matter of fact CDfR is identical to dimensional reduction to one loop order. 
In dimensional reduction the coefficients of the basic functions 
(finite, non-counterterm parts) are never projected into $n$ dimensions 
because all the algebra is performed in physical dimension of the theory 
just as in CDfR. This is not the case of DR in which $n$ can appear 
multiplying the basic functions which in turn produce different results 
from dimensional reduction.   
   
IR is a momentum space framework which somewhat resembles BPHZ in
 the sense that one algebraically manipulates the integrand of the 
amplitude in order to isolate the infinities. The idea is to isolate
 the divergences as basic divergent integrals(independent of the 
external momenta), e.g. ,    
\begin{equation}  
I_{log} (m^2) = \int_k \frac{1}{(k^2-m^2)^2}   
\end{equation}   
by using judiciously the identity   
\begin{eqnarray}  
& &\frac{1}{[(k+k_i)^2-m^2]} = \sum_{j=0}^N\frac{\left(   
-1\right) ^j\left( k_i^2+2k_i\cdot k\right) ^j}{\left(   
k^2-m^2\right) ^{j+1}}  + \nonumber \\
&+&\frac{\left( -1\right) ^{N+1}\left( k_i^2+2k_i\cdot   
k\right) ^{N+1}}{\left(   
k^2-m^2\right) ^{N+1}[ \left( k+k_i\right) ^2-m^2] },   
\label{eq:rr}   
\end{eqnarray}   
where $k_i$ are the external momenta and $N$ is chosen so that the 
last term is finite under integration over $k$. Such basic divergent 
integrals which characterise the divergent structure of the underlying 
model need not be evaluated: they can be fully absorbed in the definition 
of the renormalization constants. We shall come back to this issue in connexion
 with what is meant by MS within IR and its relation to DR and DfR. An important
 ingredient of IR is that local arbitrary counterterms parametrized by (finite)
 differences of divergent integrals of the same superficial degree of divergence
 may be cast into a set of consistency relations \cite{PRD1},\cite{PRD2}.  They 
were shown to be connected to momentum routing invariance in the loop of a Feynman
 diagram. Should they vanish (as indeed they do in DR) then one would automatically
 have momentum routing invariance and (abelian) gauge invariance. In other words by
 setting the consistency relations to zero   
(say, constrained IR (CIR)) one has the analogue to CDfR at one loop order.
 For $n=4$ they read:   
\begin{equation}  
\Upsilon_{\mu \nu}^2 \equiv \int_k \frac{g_{\mu\nu}}{k^2-m^2} -  
2\int_k \frac{k_{\mu}k_{\nu}}{(k^2-m^2)^2},   
\label{eq:CR4Q1}   
\end{equation}   
\begin{equation}  
\Upsilon_{\mu \nu}^0 \equiv \int_k \frac{g_{\mu\nu}}{(k^2-m^2)^2}-   
4\int_k\frac{k_{\mu}k_{\nu}}{(k^2-m^2)^3},   
\label{eq:CR4L1}   
\end{equation}   
\begin{equation}  
\Upsilon_{\mu \nu \alpha \beta}^2 \equiv   
g_{\{\mu\nu}g_{\alpha\beta \}}   
\int_k \frac{1}{k^2-m^2}   
-8\int_k \frac{k_{\mu}k_{\nu}k_{\alpha}k_{\beta}}{(k^2-m^2)^3},   
\label{eq:CR4Q2}   
\end{equation}   
\begin{equation}  
\Upsilon_{\mu \nu \alpha \beta}^0 \equiv   
g_{\{\mu\nu}g_{\alpha\beta \}}   
\int_k \frac{1}{(k^2-m^2)^2}   
-24\int_k \frac{k_{\mu}k_{\nu}k_{\alpha}k_{\beta}}{(k^2-m^2)^4},   
\label{eq:CR4L2}   
\end{equation}   
etc., where $g_{\{\mu\nu}g_{\alpha\beta \}}$ stands for 
$g_{\mu\nu}g_{\alpha\beta}+g_{\mu\alpha}g_{\nu\beta}+g_{\mu\beta}
g_{\nu\alpha}$.  Generically we may write $\Upsilon_{\mu \nu}^0 = 
\alpha_i g_{\mu \nu}$, etc. with $\alpha_i$ arbitrary and finite.  
   
It is well known, however, that a shift in $k$ is immaterial only 
if $\Delta_s \le 0$,
$\Delta_s$ being the superficial degree of divergence, otherwise  
a  surface term should be added. This is an indication that one 
should be careful in what concerns the momentum routing when 
divergences higher than logarithmic arise in Feynman diagram 
calculations. Perturbation theory makes a peculiar use of this 
feature for in some cases gauge invariance relies on adopting a 
special momentum routing \cite{JACKIWCA}. A related issue is that
 whilst a shift in the integration variable is allowed within dimensional 
regularization, the algebraic properties of $\gamma_5$ clash with analytical
 continuation on the space-time dimension.  In such cases, in IR we work with 
arbitrary values for the  consistency relations until  the end of the calculation
 so that physical conditions determine (or not) their value. For instance, a 
democratic display of the Adler-Bardeen-Bell-Jackiw triangle anomaly can only 
be achieved for arbitrary values of $(\ref{eq:CR4L1})$.    
   
\subsection{Examples}   
\label{sec:examples}  
   
Here we illustrate the correspondence between the different regularisation
 frameworks in the context of a MS renormalisation scheme in massless $\phi^4$-theory
 and $QED$. In particular we study the Ward identity involving the QED vertex function
whose finite parte is easily obtained within IR and analyse the role played by the 
consistency relations and arbitrary momentum routing. The $4$-point function of the 
$g/4! \phi^4_4$-theory to one loop order $\Gamma^{4}_\hbar(p)$is proportional to 
($\mu$ is an infrared cutoff)    
\begin{equation}  
{\cal{A}} \equiv \int_k \frac{1}{(k^2-\mu)^2[(k-p)^2-\mu^2]}\, .   
\end{equation}   
In IR we apply $(\ref{eq:rr})$ once to get,    
\begin{equation}  
{\cal{A}} = 3 I_{log}(\mu^2) - b \int_0^1 dz \ln \Bigg( \frac{p^2z(z-1)}{-\mu^2}+1 \Bigg) \, ,
\label{eq:4pf}   
\end{equation}   
with $p^2=s,t,u$. In $(\ref{eq:4pf})$ we must separate the ultraviolet and infrared 
divergences (for the case $\mu \rightarrow 0$) before proceeding to renormalisation. 
This can be easily accomplished by using the identity   
\begin{equation}  
I_{log}(\mu^2)  = I_{log}(\lambda^2) + b \ln \Big( \frac{\lambda^2}{\mu^2} \Big)\, ,   
\label{eq:scarel}   
\end{equation}   
which holds for arbitrary $\lambda $. This enables us to write   
\begin{eqnarray}   
\Gamma^4_\hbar (p) &=& \frac{g^2}{2}  \Bigg( I_{log} (\lambda^2) + b \ln 
\Big( \frac{\lambda^2}{\mu^2}\Big)-\nonumber \\ &-& b \int_0^1 dz \ln 
\Bigg( \frac{p^2z(z-1)}{-\mu^2}+1 \Bigg) \Bigg|_{p^2=s} \Bigg) \nonumber \\   
&=& \frac{g^2}{2}  \Bigg( I_{log} (\lambda^2) + b \ln \Big( \frac{\lambda^2}
{\mu^2}\Big)- \nonumber \\ &-& b (-2 + \alpha_s \ln \Big( \frac{\alpha_s+1}
{\alpha_s-1}\Big)\Bigg)\, ,   
\end{eqnarray}   
with $\alpha_s = \sqrt{4 \mu^2/s +1}$. Hence we define the MS within the 
IR method by subtracting $I_{log} (\lambda^2)$ to yield   
\begin{equation}  
\Gamma^{4 R}_\hbar \Big|_{MS} (p) = \frac{b \,\,g^2}{2}  \Bigg( \ln \Big( 
\frac{{\bar{\lambda}}^2}{\mu^2}\Big)- \alpha_s \ln 
\Big( \frac{\alpha_s+1}{\alpha_s-1}\Big)\Bigg)\, .   
\end{equation}   
which is just the result obtained in DfR \cite{PV3A} with ${\bar{\lambda}}^2 
\equiv \lambda^2 \mbox{e}^2 = \bar{M}^2$. Notice that in the limit where $\mu 
\rightarrow 0$, $\mu$ cancels out in the  equation above, as it should . That 
is because the infrared divergent piece of the  logarithm $ b \ln \Big( 
\frac{\lambda^2}{\mu^2}\Big)$ cancels out with another piece coming from 
the finite part of the amplitude. this enables us to write  
\begin{equation} 
{\cal{A}}^R_{MS} = b \ln \Big( \frac{\bar{\lambda}^2}{p^2} \Big) \, .  
\end{equation}  
Therefore we have in the MS scheme as defined above for IR the same prescription 
as defined by equation $(\ref{eq:ft})$ in DfR. Moreover one can verify that
 ${\bar{\lambda}}$ plays the role of renormalization group scale in the 
Callan-Symanzik equation. This is expected since it parametrises the 
arbitrariness in separating the divergent from the finite part. In the 
massless limit the Callan-Symanzik equation for the $4$-point function reads   
\begin{equation}  
\Bigg( {\bar{\lambda}} \frac{\partial}{\partial {\bar{\lambda}}}   
+ \beta \frac{\partial}{\partial g} + 4 \gamma_\phi \Bigg)
\Gamma^{(4)}_R (p^2) = 0 \, ,   
\end{equation}   
from which we compute the standard value $\beta = 3 g^2/(16 \pi^2)$.   
   
We also expect IR to be identical to {\it{dimensional reduction}} 
(as DfR is) to one loop level for the Lorentz algebra which determines 
the coefficients of the finite parts in IR is effected in the integer 
dimension, say $n=4$. In order to illustrate this point, as well as to 
pinpoint the role played by the consistency relations in IR  in connection 
with momentum routing and gauge invariance let us study the QED Ward identity
 involving the vertex function in IR. The electron self-energy in the Feynman 
gauge is written as ($e^2$ = 1)   
\begin{eqnarray}  
\Sigma &=& \int_k \gamma_\mu \frac{1}{\not{k}+\not{k}_1-m}\gamma_\nu 
\frac{g_{\mu \nu}}{(k+k_2)^2-\mu_\gamma^2} \nonumber \\ &\Rightarrow& -
\frac{\Sigma}{2} = (\not{k}_1 - 2m) I + \gamma_\mu I^{\mu}   
\end{eqnarray}    
where $\mu_\gamma$  is an infrared regulator, $k_1$,$k_2$ are arbitrary
 momenta running in the loop such that $k_1 - k_2 = p$, $p$ being the 
external momentum, which we shall parametrise by $\alpha$ by setting 
$k_1 = (1+\alpha) p$ and $k_2 = \alpha p$), and    
$$   
I, I^\mu = \int_k \frac{1, k^\mu}{[(k+k_1)^2-m^2][(k+k_2)^2-\mu_\gamma^2]} \, .   
$$   
Now within the spirit of IR we use $(\ref{eq:rr})$ to write   
\begin{eqnarray}   
I &=& I_{log}(m^2) - \int_k \frac{k_2^2+2 k \cdot k_2 + m^2 - 
\mu_\gamma^2}{(k^2-m^2)^2[(k+k_2)^2-\mu_\gamma^2]} - \nonumber \\
 &-& \int_k \frac{k_1^2+2 k \cdot k_1}{(k^2+m^2)[(k+k_2)^2-
\mu_\gamma^2][(k+k_1)^2 - m^2]} \nonumber \\   
&=& I_{log}(m^2) - b \,\, Z_0 (\mu_\gamma^2,m^2,p^2;m^2) \, ,   
\end{eqnarray}   
in which $Z_0$ is part of a class of  functions which characterise
 Feynman diagram calculations to one loop order,   
\begin{eqnarray}  
&& Z_k (\lambda_1^2, \lambda_2^2, p^2; \lambda^2) = \nonumber \\ 
&& \int_0^1 dz \,\, z^k \ln \frac{p^2 z (1-z) + (\lambda_1^2-
\lambda_2^2)z - \lambda_1^2}{-\lambda^2}.   
\label{eq:zks}  
\end{eqnarray}   
We can similarly calculate $I_\mu$ to get   
\begin{eqnarray}   
& & I^\mu = -\frac{1}{2}(k_1+k_2)^\mu I_{log} (m^2) + 
(k_1+k_2) \, \lambda_1 \, p^\mu    
\nonumber \\   
&+& b \, k_2^\mu \, Z_0(\mu_\gamma^2,m^2,p^2;m^2)  - b \,
 p^\mu \, Z_1 (\mu_\gamma^2,m^2,p^2;m^2).   
\end{eqnarray}    
where $\lambda_1$  is defined from $(\ref{eq:CR4L1})$ as    
$$   
\Upsilon^0_{\mu \nu} \equiv \lambda_1 g_{\mu \nu}   
$$   
and is in principle undetermined. The limit $\mu_\gamma 
\rightarrow 0$ is well defined for the $Z_k$ functions and 
$Z_1 (0,m^2,p^2;m^2)= 1/2 Z_0 (0,m^2,p^2;m^2)$. This enables us to write   
\begin{eqnarray}  
\Sigma (p) &=& - (\not{p} - 4 m) \Big( I_{log}(m^2) - b \, 
Z_0 (0,m^2,p^2;m^2)\Big)\nonumber \\
&+& (2 \alpha + 1) \lambda_1 \not{p}.   
\label{eq:ese}   
\end{eqnarray}   
   
In order to establish the value of $\lambda_1$ it is natural to check 
whether the  Ward identity which relates $\Sigma$ to the vertex function 
can place any constraint on $\lambda_1$.    
Consider the QED vertex function with incoming momenta $p$ and $q$ and
 outgoing momentum $p+q$ \footnote{Because the QED vertex function is
 superficially logarithmically divergent, it must be momentum routing 
independent.}:   
\begin{equation}  
- i \Lambda^\mu (p,q) = i \int_k \gamma_\alpha 
\frac{g^{\alpha \beta}}{k^2-\mu_\gamma^2}   
\gamma_{\beta}\frac{1}{\not{k} + \not{q} -m}
\gamma^{\mu}\frac{1}{\not{k} - \not{p}-m}   
\end{equation}   
Within the framework of IR we can write, after some 
tedious yet straightforward algebra,    
\begin{eqnarray}   
& &- i \Lambda^\mu (p,q) = \gamma_\nu \Upsilon_0^{\mu \nu} + 
\gamma^\mu I_{log} (m^2) +  2 b \gamma^\mu (\widetilde{Z}-Z_0) + 
\nonumber \\ &+& b \xi^{00} \, \Big( 4 m (p + q)^\mu -2 m^2 \gamma^\mu 
- 2 \not{q}\gamma^\mu \not{p}\Big)+\nonumber \\   
&+& 2 b ( \not{p} + \not{q})\gamma^\mu \not{p}\, \xi^{01} + 2 b 
\not{q}\gamma^\mu(\not{p} + \not{q})\, \xi^{10} - \nonumber \\ 
&-& 8 m b (p^\mu \, \xi^{01} + q^\mu \, \xi^{10}) -4 b (p^\mu \not{p} 
\xi^{02} + q^\mu  \not{q} \xi^{20}) + \nonumber \\   
&+& b (p^\mu \not{q} + q^\mu  \not{p} ) \, \xi^{11} \, ,   
\label{eq:vf}   
\end{eqnarray}   
where    
\begin{eqnarray}   
&& Z_0 = Z_0 (m^2,m^2,(p-q)^2;m^2)\, , \nonumber \\   
&& \widetilde{Z} = \widetilde{Z}(\mu_\gamma^2,m^2,p,q) \equiv 
\nonumber \\ && \int_0^1 dz\int_0^{1-z}dy \,\,   
\ln \Bigg( \frac{Q(p,q,y,z,\mu_\gamma^2,m^2)}{-m^2}\Bigg) \, ,\nonumber \\   
&& \xi^{mn} = \xi^{mn}(\mu_\gamma^2,m^2,p,q) \equiv \nonumber \\ 
&& \int_0^1 dz\int_0^{1-z}dy \,\, \frac{y^n z^m}
{Q(p,q,y,z,\mu_\gamma^2,m^2)}\, , \nonumber \\   
&& Q(p,q,y,z,\mu_\gamma^2,m^2) = p^2 y(1-y) + q^2 z(1-z) - \nonumber \\  
&& - 2 p\cdot q \, y z   +  (z+y)(\mu_\gamma^2-m^2)-\mu_\gamma^2 \, .   
\end{eqnarray}   
Now we write $\Upsilon_0^{\mu \nu}= \lambda_2 g^{\mu \nu}$ in 
$(\ref{eq:vf})$, $\lambda_2$ being an arbitrary parameter, and redefine   
\begin{eqnarray}   
\Sigma (\not{p}) &=& \widetilde{\Sigma} (\not{p}) + (2 \alpha +1)
\lambda_1 \not{p} \, , \nonumber \\   
\Lambda^\mu (p,q) &=& \widetilde{\Lambda}^\mu (p,q) + \gamma^\mu \lambda_2   
\end{eqnarray}   
which with the help of the relations displayed in the appendix, 
enable us to  verify promptly that   
\begin{equation}  
(p-q)_\mu {\widetilde{\Lambda}}^\mu (p,q) = \widetilde{\Sigma} 
(\not{p}) - \widetilde{\Sigma} (\not{q}) \, .   
\end{equation}   
Hence the Ward identity is fulfilled if   
$$   
\lambda_2 = (2 \alpha +1) \lambda_1.   
$$   
The natural choice is to set $\lambda_1 = \lambda_2 = 0$, which automatically 
implements both gauge and momentum routing invariance (CIR). Notice that by 
setting $\lambda_1 = 0$ in  $(\ref{eq:ese})$ leads to the same result as obtained
 in CDfR and dimensional reduction   
\cite{PV4} (which is however different from DR within the same subtraction scheme)
 as we too have worked in four dimensions.    
   
This illustrates the equivalence between CIR, CDfR and dimensional reduction
 to one loop level. For instance, the superfield calculation of the one loop 
correction to the vector propagator is gauge invariant in dimensional 
reduction scheme only if \cite{GRISARU}    
\begin{equation}  
{\cal{I}}^\mu = \int_k \frac{p^\mu+2k^\mu}{k^2(k+p)^2}=0\, .   
\end{equation}   
We can easily evaluate the integral above within IR to show that
 it reduces to   
\begin{equation}  
{\cal{I}}^\mu _{IR} = p^\nu \Upsilon_{\mu \nu}^0 \,   
\end{equation}   
showing that CIR ($\Upsilon_{\mu \nu}^0 = 0$) may be safe framework
 to handle the problem. In principle IR is  generalisable to higher
 loop calculations avoiding breakdown of symmetries such as gauge 
invariance and supersymmetry \cite{WIP}.    
   
\section{IR, BPHZ and DfR: A Two Loop Example}   
   
\begin{figure}[h]  
\bc  
\centerline{  
\epsfxsize=4in  
\epsffile{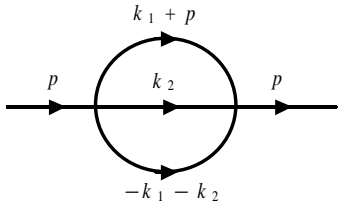}  
}  
\vskip-2.0cm
\caption[setsun]{\label{setsun} Sunset Diagram}  
\ec  
\end{figure}

\begin{figure}[h]  
\bc  
\centerline{  
\epsfxsize=4in  
\epsffile{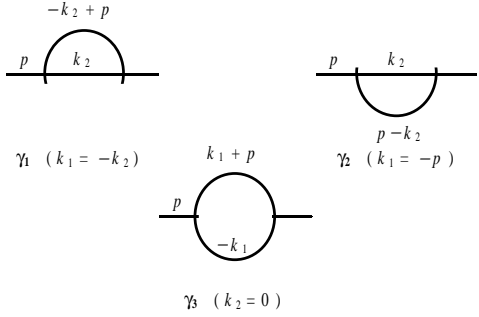}  
}  
\vskip-1.0cm
\caption[subgraphs]{\label{subgraphs} Subgraphs $\gamma_1, \gamma_2$ and $\gamma_3$}  
\ec  
\end{figure}

In order to illustrate the correspondence between the BPHZ formalism 
and IR, we shall compute the $2$-loop correction to the $2$-point 
function in $\phi^4_4$ theory  in both methods and compare with DfR. 
The amplitude  is depicted in figure $(\ref{setsun})$ (setting-sun diagram). 
The computation of the finite part of this diagram is notoriously difficult 
in DR, for instance. However for both IR and DfR \cite{PV3B} it can be 
readily displayed.    
 The BPHZ scheme relies on the forest formula to perform the 
subtraction of the divergences from an amplitude \cite{BPHZ}. Let 
$I_G^\infty$ be the integrand of such amplitude associated with   
a graph $G$. Then the subtracted integrand is given by   
\begin{equation}  
R_G = \sum_{U \in \phi } \prod_{\gamma \in U} - t^{d(\gamma)}_\gamma I_G^\infty  
\label{eq:forform}   
\end{equation}   
where $\phi$ is the set of all the forests $U$ of $G$, including the 
empty set \footnote{A forest $U$ of a diagrama $G$ is the set of all 
subgraphs of G, including $G$ itself, which are neither overlapping 
nor trivial.}, $d(\gamma)$ is the  superficial degree of divergence 
of the subgraph $\gamma$ and $t^{d(\gamma)}$ is the Taylor operator 
which corresponds to an expansion around $0$ to order $d(\gamma)$ in 
the external momentum to the subgraph. For the sunset diagram the subgraphs 
are shown in figure $(\ref{subgraphs})$.   
Therefore we can write $\phi = \{\emptyset , G , \gamma_1, 
\gamma_2, \gamma_3, G \gamma_1, G \gamma_2, G \gamma_3 \} $. Thus   
\begin{eqnarray}  
R_G &=& (1-t^{0}_{{\gamma}_1} -t^{0}_{{\gamma}_2} -t^{0}_{{\gamma}_3} 
-t^{2}_G + \nonumber \\
&+& t^{2}_G t^{0}_{{\gamma}_1} + t^{2}_G t^{0}_{{\gamma}_2} + 
t^{G} t^{0}_{{\gamma}_3}) I_G^\infty \label{eq:bphzsub}   
\end{eqnarray}   
which are ordered so that if $\gamma_1 \subset \gamma_2$ then 
$t_{\gamma_{1}}$ lies on the right of  $t_{\gamma_{2}}$. The 
amplitude for the sunset diagram is superficially quadratically 
divergent. It reads   
\begin{eqnarray}   
\Gamma^{(2)}_{\hbar^2} (p) &=& \frac{g^2}{6} \int_{k_1,k_2} 
I_G^\infty \, , \nonumber \\   
I_G^\infty &=& \Delta (k_1+p) \Delta (k_2) 
\Delta (k_1 + k_2) \,\,\, {\mbox{with}}  \nonumber \\   
\Delta (k) &\equiv& \frac{1}{k^2 - \mu^2} \, .   
\label{eq:ampss}  
\end{eqnarray}   
Therefore we get, using $(\ref{eq:bphzsub})$,   
\begin{equation}  
R_G = I_G^\infty - t_G^2 I_G^\infty - \Delta(k_2)^2 ( 1 - t_G^2  )\Delta (k_1 +p)\, .    
\end{equation}   
If one uses a regulator, it can be shown that the term $\Delta(k_2)^2 ( 1 - t_G^2  )
\Delta (k_1 +p)$ above actually vanishes. This comes from the fact that    
$$   
F(p) = \int_k (1 - t_G^2) \Delta (k+p)   
$$   
is independent of $p$ and therefore $F(p)=F(0)=0$. Thus according to the 
BPHZ forest formula, the finite amplitude is obtained through the subtractions:  
\begin{eqnarray}   
&& \int_{k_1,k_2} R_G = \int_{k_1,k_2} I_G^\infty - 
\int_{k_1,k_2} t_G^2 I_G^\infty \, ,\nonumber\\&&\int_{k_1,k_2} 
t_G^2 I_G^\infty  = I_{111} (\mu^2) - p^2 I_{112} (\mu^2)  + \nonumber \\  
&& +  4 \int_{k_1,k_2} (p \cdot k_1)^2 \Delta(k_1)^3\Delta(k_2)\Delta(k_1+k_2) \,\, ,  
\label{eq:bphzss}   
\end{eqnarray}   
where $I_{mnp}(\mu^2)$ stands for  
\begin{equation} 
\int_{k_1,k_2} \frac{1}{(k_1^2-\mu^2)^m(k_2^2-\mu^2)^n((k_1+k_2)^2-\mu^2)^p}\, .  
\label{eq:imnp}  
\end{equation}  
The integrals $I_{mnp}$ are  convergent  only if $m+p>2$,  $n+p>2$, $m+n>2$, 
$m+n+p>4$ are satisfied \footnote{It will also be useful to know that 
$I_{mnp}^{ijk}=\int_{k_1,k_2} \frac{(k_1^2)^i (k_2^2)^j (k_1\cdot k_2)^k}
{(k_1^2-\mu^2)^m(k_2^2-\mu^2)^n((k_1+k_2)^2-\mu^2)^p}$ converges if 
$2 m+2 p>4+2i+k$,  $2 n+2 p>4+2j+k$, $m+n+p>i+j+k+4$ and $m+n > i+j+k+2$ 
are satisfied. }. In order to display the finite part of the setting sun 
amplitude there is still some work to do in $(\ref{eq:bphzss})$. In particular
 one would have to adopt a regulator to proceed in this task, the result being 
regulator independent as guaranteed by the BPHZ scheme. That is to say, the BPHZ 
scheme guarantees to us is which subtractions one ought to perform  in order to 
render the amplitude finite and that the result is regularization independent.   
  
Now let us evaluate $(\ref{eq:ampss})$ in the light of IR. Because $d(G)=2$ 
all we need is to apply $(\ref{eq:rr})$ up to $N=2$ in $\Delta (k_1+p)$ so to 
display the infinities in terms of basic divergent integrals which are cleared 
out of external momenta to get  
\begin{eqnarray}  
\Delta (k_1+p) &=& \Delta (k_1) - (p^2+2 k_1 \cdot p)\Delta^2 (k_1) \nonumber \\  
&+& (p^2 + 2 p \cdot k_1)^2 \Delta^3(k_1) \nonumber \\ 
&-& (p^2 + 2 p \cdot k_1)^3 \Delta^3(k_1) \Delta (k_1+p) \, ,  
\end{eqnarray}   
which enables us to write  
\begin{eqnarray}  
& &\frac{6}{g^2}\Gamma^{(2)} (p) = I_{111} (\mu^2) - p^2 
I_{112} (\mu^2)  \nonumber \\  
&+&  4 \int_{k_1,k_2} (p \cdot k_1)^2 \Delta(k_1)^3\Delta(k_2)
\Delta(k_1+k_2) + p^4  I_{311}
\nonumber \\  
&-& \int_{k_1,k_2} (p^2 + 2 p\cdot k_1)^3 
\Delta(k_1)^3\Delta(k_2) \times \nonumber \\
&\times& \Delta(k_1+k_2)\Delta(k_1+p) \, .\label{eq:sperou}  
\end{eqnarray}  
Notice that the first three terms in the rhs of the equation above 
are just the terms which were prescribed to be subtracted using the 
BPHZ scheme; the fourth term and fifth term (let us call them $F_4$ 
and $F_5$ for definiteness ) are clearly  divergent but   
their difference is finite. To see that let us isolate the divergence 
in both terms as a function of one loop momentum variable only using 
$(\ref{eq:rr})$. Thus  
\begin{eqnarray} 
&& F_4 = p^4 I_{log}(\mu^2) \int_{k_1}\Delta(k_1)^3  - p^4 \times \nonumber \\ &&
\times \int_{k_1,k_2} (k_1^2+2 k_1   
\cdot k_2) \Delta(k_1)^3 \Delta(k_2)^2 \Delta(k_1+k_2) \, ,  
\end{eqnarray}  
whereas  
\begin{eqnarray}  
&& F_5 =  I_{log}(\mu^2) \int_{k_1} (p^2+ 2 p \cdot k_1)^3 
\Delta(k_1)^3\Delta(k_1+p) \nonumber \\  
&&- \int_{k_1,k_2} (p^2+2 k_1 \cdot p)^3(k_1^2+2k_1\cdot k_2)
\Delta(k_1)^3 \times \nonumber \\
&& \times \Delta(k_2)^2\Delta(k_1+k_2)\Delta(k_1+p) \,\,\, .  
\end{eqnarray}  
By using Feynman parameters to evaluate the first terms in $F_4$
and $F_5$ one can show that they cancel out. Therefore we can write 
the finite part ${\cal{F}}= F_4 - F_5$ of the setting-sun amplitude as  
\begin{eqnarray}  
&& {\cal{F}} = \frac{g^2}{6} \int_{k_1,k_2} \Big(  
(p^2+2 k_1 \cdot p)^3(k_1^2+2k_1\cdot k_2) \times \nonumber \\ && \times 
\Delta(k_1)^3\Delta(k_2)^2 \Delta(k_1+k_2) \Delta(k_1+p) - \nonumber \\  
 && - p^4 (k_1^2+2 k_1 \cdot k_2)\Delta(k_1)^3\Delta(k_2)^2\Delta(k_1+k_2) \Big) .  
\end{eqnarray}  
In order to make  contact with other regularisation/renormalisation frameworks
 let us take a closer look at the divergent structure of $\Gamma^{(2)}(p)$, namely  
\begin{eqnarray} 
& & \frac{g^2}{6} \Big(  I_{111}(\mu^2) - p^2 I_{112}(\mu^2)+ \nonumber \\ &+&  
4 \int_{k_1,k_2} (p \cdot k_1)^2 \Delta(k_1)^3\Delta(k_2)\Delta(k_1+k_2) \Big).  
\end{eqnarray}  
It is easy to show that the third term above may be written as  
\begin{eqnarray}  
&& p^2 (I_{112}(\mu^2) + \mu^2 I_{113}(\mu^2)) \,\,\,\,\, \mbox{and that} \nonumber \\  
&& I_{113} = -\frac{b}{2 \mu^2} I_{log}(\mu^2) + N  \,\,\,\,\  , \, N \,\,\, \mbox{finite}   
\nonumber \\  
&& N = - \int_{k_1,k_2}  (k_1^2+2 k_1   \cdot k_2) \nonumber \\
&& \times \Delta(k_1)^3\Delta(k_2)^2\Delta(k_1+k_2)\,\, .  
\end{eqnarray}  
Also let us split the logarithmic divergence using $(\ref{eq:scarel})$.   
Collecting all the results together we have \footnote{In CDfR the
 dimensionful constants are taken from the logarithmic divergent 
pieces only \cite{PV3} }  
\begin{eqnarray}  
&& \Gamma^{(2) R}_{\hbar^2} (p) = \frac{g^2}{6} \Bigg(   
- \frac{b^2 \, p^2}{2} \ln \Big( \frac{\lambda^2}{\mu^2} \Big) \nonumber \\ 
&& -  \int_{k_1,k_2}   (p^4 + \mu^2 p^2)(k_1^2+2 k_1   
\cdot k_2) \Delta(k_1)^3\Delta(k_2)^2 \times \nonumber \\ && \times \Delta(k_1+k_2)    
 + \int_{k_1,k_2}  (p^2+2 k_1 \cdot p)^3(k_1^2+2k_1\cdot k_2) \times \nonumber \\
&& \times \Delta(k_1)^3\Delta(k_2)^2\Delta(k_1+k_2)\Delta(k_1+p)  
 \Bigg)   \,\, ,
\label{eq:pss}  
\end{eqnarray}  
where   
\begin{equation} 
 \Gamma^{(2)}_{\hbar^2} (p) - \Gamma^{(2)R}_{\hbar^2} (p) = 
 \frac{g^2}{6} ( I_{111}(\mu^2) - \frac{b p^2}{2} I_{log}(\lambda^2)) \, .  
\label{eq:divss}  
\end{equation}  
The equation $(\ref{eq:pss})$ obtained within IR is the momentum 
space analogue of the result presented in \cite{PV3B} in DfR,  
\begin{eqnarray}  
&& \Gamma_{\hbar^2}^{(2)}\Bigg|_R(x,M) = \frac{g^2}{96(4 \pi^2)^3} 
\Big[ (\square - 9\mu^2) (\square-\mu^2)  \nonumber \\ && \Big( \mu^2 
K_0(\mu x) K_1^2(\mu x)+ \mu^2 K_0^3(\mu x)\Big)   
\nonumber \\  && + 2 \pi^2 \ln \frac{\bar{M}^2}{\mu^2} 
(\square - \mu^2) \delta^{(4)} (x)  
\Big] \, .  
\end{eqnarray}  
Here too, as it was claimed in \cite{PV3B} for DfR, one has been
 able to display the finite part of the setting sun diagram in a 
closed and compact form in an easy fashion with the advantage of working 
directly in the momentum space. Notice that the scale $\lambda$ in 
($\ref{eq:pss}$) is just the analogue of $\bar{M}$ in the equation 
above and plays the role of scale in the Callan-Symanzik equation satisfied 
by $\Gamma^{(2)}_R (p)$. For instance, for $\mu \rightarrow 0$, 
$\Gamma^{(2)}_R (p^2)$ is well defined and obbeys   
\begin{equation}  
\Bigg( {{\lambda}} \frac{\partial}{\partial {{\lambda}}}   
+ \beta \frac{\partial}{\partial g} + 2 \gamma_\phi \Bigg)\Gamma^{(2)}_R (p^2) = 0 \, ,   
\end{equation}   
from which we may calculate the lowest order value of $\gamma_\phi$, namely  
\begin{equation} 
\gamma_\phi = \frac{1}{12} \frac{g^2}{(16 \pi^2)^2}\, .  
\label{eq:gp}  
\end{equation}  
In fact it can be shown that $\gamma_\phi$ to lowest order is simply the 
coefficient of the logarithmic divergence $\ln \Lambda^2/\mu^2$ \cite{PESKIN}.
 By using a general parametrisation for $I_{log}(\lambda^2)$ \cite{PRD2}, 
viz. \footnote{Such parametrisation is constructed based on 
 $\del I_{log}(\lambda^2)/\del \lambda^2 = -b/\lambda^2$.}   
\begin{equation} 
I_{log}(\lambda^2) = b \ln \Big( \frac{\Lambda^2}{\lambda^2} \Big) + \eta \,   
\end{equation}  
($\Lambda$ is an UV cutoff and $\eta$ an arbitrary constant) in 
$(\ref{eq:divss})$ we see that the coefficient of the logarithmic 
divergence evaluates to $\gamma_\phi$ given in $(\ref{eq:gp})$. Alternatively 
one can apply DR to evaluate $I_{log}(\lambda^2)$ which gives
 $b \Gamma (\epsilon)= b (1/\epsilon - \gamma_E + O(\epsilon) )$.  
  
We can also study the setting sun diagram with arbitrary routing . Let 
us split the external momentum $p$ between the upper and lower lines
 in figure $(\ref{setsun})$ so that the amplitude reads  
\begin{eqnarray}  
\Gamma^{(2)}_{\hbar^2}(p , \alpha , \beta)&=& \frac{g^2}{6}\int_{k_1,k_2} 
\Delta (k_1+ \alpha p) \Delta (k_1 + k_2) \Delta (k_2 + \beta p) \nonumber \\   
&\equiv& \frac{g^2}{6}  \int_{k_1,k_2} I_G^\infty (\alpha, \beta ) \, .  
\label{eq:arbrou}  
\end{eqnarray}  
with $\alpha + \beta =1$. Using the forest formula one can show 
within the BPHZ framework that  
\begin{eqnarray}  
&& \int_{k_1,k_2} R_G = \int_{k_1,k_2} I_G^\infty (\alpha, \beta ) -
 I_{111}(\mu^2) + (\alpha + \beta)^2 I_{112} (\mu^2)\nonumber \\  
&& - 4 (\alpha + \beta)^2 \int_{k_1,k_2} (p \cdot k_1)^2 
\Delta(k_1)^3\Delta(k_2)\Delta(k_1 + k_2) \,\, .  
\end{eqnarray}  
To work out $(\ref{eq:arbrou})$ within IR, it suffices to expand
 $\Delta (k_1 + \alpha p)$ and $\Delta (k_2 + \beta p)$ using 
$(\ref{eq:rr})$ up to $N=2$ to get  
\begin{eqnarray}  
&& \Gamma^{(2)}_{\hbar^2}(p , \alpha , \beta) =   
I_{111}(\mu^2) - (\alpha + \beta)^2 p^2 I_{112} (\mu^2)\nonumber \\  
&&+ 4 (\alpha + \beta)^2  \int_{k_1,k_2} (p \cdot k_1)^2 
\Delta(k_1)^3\Delta(k_2)\Delta(k_1+k_2)\nonumber \\  
&&+ \,\,\,  finite \,\, (\alpha,\beta = 1-\alpha).  
\end{eqnarray}  
which turns out to be identical to $(\ref{eq:sperou})$, as it 
should, because the finite part above can be shown to be independent of $\alpha$.
 No consistency relations have appeared in this example as the momentum routing 
independence in the setting sun diagram is trivial. In \cite{OAMC} we calculate 
the $\beta$ function of $\phi^4_4$ theory to two loop order within IR.  
  
\section{Gluon Self-Energy of QCD}  
  
In both abelian and nonabelian gauge field theory the gauge boson self energy 
is bound by gauge invariance to have the structure $i (p^2 g_{\mu \nu} - 
p_\mu p_\nu) \Pi (p^2)$. For $QCD$, the cancellations that lead to this structure 
are more complex than for $QED$ and relies on a gauge invariant regularisation 
framework to handle both the ultraviolet and infrared divergences. It is well 
known that adding a small mass for the gluon breaks gauge invariance although it 
may be safely done for the photon.   
  
It will be interesting to see how gauge invariance is implemented in IR for the 
gluon self energy in connection with the consistency conditions $(\ref{eq:CR4Q1})$
-$(\ref{eq:CR4L2})$.  
Let $p$ be the external momentum. The diagrams that contribute to the gluon self 
energy to one loop order are well known: 1) the gluon tadpole $\Pi_{\mu \nu}^{ab}(1)$,
 2)the gluon loop $\Pi_{\mu \nu}^{ab}(2)$, 3) the ghost loop  $\Pi_{\mu \nu}^{ab}(3)$ 
and 4) the fermion loop $\Pi_{\mu \nu}^{ab}(4)$. Hence the gluon self energy is given by  
\begin{equation}   
\Pi^{ab}_{\mu \nu}=\Pi^{ab}_{\mu \nu}(1)+\Pi^{ab}_{\mu \nu}(2)+
\Pi^{ab}_{\mu \nu}(3)+\Pi^{ab}_{\mu \nu}(4)    
\end{equation}    
The Feynman rules in momentum space for QCD can be found in any textbook 
\cite{PESKIN}.  It is straightforward to show that  
\begin{eqnarray}   
\Pi^{ab}_{\mu \nu}(1) &=& -g^2C_2(G) \delta^{ab}3\int_k \frac{g_{\mu \nu}}
{k^2-\mu^2}\nonumber \\
&=&    -3 g^2 g_{\mu \nu} C_2(G) \delta^{ab} I_{quad}(\mu^2)    
\end{eqnarray}    
The gluon loop is equally simple to be displayed within IR. It reads  
\begin{equation}   
\Pi^{ab}_{\mu \nu}(2)= \frac{1}{2} \int_k  g^2 f^{acd} f^{bcd} N_{\mu \nu}  
\frac{-i}{k^2-\mu^2}\frac{-i}{(k+p)^2-\mu^2} \,\, ,    
\end{equation}    
where   
\begin{eqnarray}    
N^{\mu \nu}&=&[g^{\mu \rho}(p-k)^{\sigma}+g^{\rho \sigma}(2k+p)^{\mu} + 
g^{\sigma \mu}(-k-2p)^{\rho}]  \nonumber \\  
&\times&  
[\delta^{\nu}_\rho (k-p)_{\sigma}+g_{\rho \sigma}(-2k-p)^{\nu}+
\delta^{\nu}_{\sigma}(k+2p)_\rho]  \nonumber \\    
&=& 2p_\mu p_\nu -5(p_\mu k_\nu +p_\nu k_\mu) -10 k_\mu k_\nu \nonumber \\
&-& g_{\mu \nu}[(p-k)^2+ (k+2p)^2],    
\end{eqnarray}    
Using that  
\begin{equation}   
(p-k)^2+ (k+2p)^2=(k+p)^2+k^2+4p^2 ,  
\end{equation}    
we may write  
\begin{eqnarray}    
\Pi^{ab}_{\mu \nu}(2)&=& -\frac 12 g^2 C_2(G) \delta^{ab} [(2p_\mu p_\nu - 
4p^2 g_{\mu \nu})J (p^2,\mu^2)  \nonumber \\ 
&-& g_{\mu \nu}(2 I_{quad}(\mu^2) + p^\alpha p^\beta 
\Upsilon_{\alpha \beta}^0)   \nonumber \\    
&-& 10(\, p_\nu J_\mu (p^2,\mu^2) + J_{\mu \nu}(p^2,\mu^2) \, )] \,  ,  
\end{eqnarray}    
whereas the ghost loop is given by  
\begin{eqnarray}    
&& \Pi^{ab}_{\mu \nu}(3)= -\int_k \frac{i}{k^2-\mu^2}
\frac {i}{[(k+p)^2-\mu^2]}g^2f^{dac}f^{cbd} \times
\nonumber \\ && \times (p+k)_\mu k_\nu 
= - g^2 \delta^{ab} C_2(G) (p_\nu J_\mu  (p^2,\mu^2) 
\nonumber \\ && + J_{\mu \nu}  (p^2,\mu^2) ) \,\, ,   
\end{eqnarray}    
in which we have used the notation  
\begin{eqnarray}  
J_{\mu \nu}&=&\Theta^{(2)}_{\mu \nu}-  
 p^2 \Theta^{(0)}_{\mu \nu}    
+ 4 p^\alpha p^\beta \Theta^{(0)} _{\mu \nu \alpha \beta}    \\    
&+& b \Bigg[\frac {p_\mu p_\nu}{3} \Big[\frac 16 -\frac 1{p^2}(p^2-\mu^2)Z_0\Big]\\    
&-& \frac{p^2 g_{\mu \nu}}{6}\Big[\frac 13 +\frac 1{2p^2}(-p^2+4m^2)Z_0\Big] \Bigg] \\  
J_\mu &=& -2p^\alpha \Theta_{\alpha \mu}^{(0)} + \frac 12 \, p_\mu \, b  \, Z_0  \\  
J&=& I_{log}(\mu^2) - Z_0\\  
\end{eqnarray}  
with the $Z_0$ functions defined as in $(\ref{eq:zks})$  
\begin{equation} 
Z_0 = Z_0 (\mu^2,\mu^2,p^2;\mu^2)   
\end{equation}  
and  
\begin{eqnarray}   
\Theta_{\mu \nu}^{(0)}&=&\int_k  \frac{k_\mu k_\nu}{(k^2-\mu^2)^3}  \nonumber \\  
\Theta^{(2)}_{\mu \nu}&=& \int_k \frac{k_\mu k_\nu}{(k^2-\mu^2)^2}  \nonumber \\  
\Theta_{\mu \nu \alpha \beta}^{(0)} &=&    
\int_k \frac{k_\mu k_\nu k_\alpha k_\beta}{(k^2-\mu^2)^4}  \, .  
\end{eqnarray}  
Hence we can write $(\ref{eq:CR4L1})$ as $\Upsilon_{\mu \nu}^0 = g_{\mu \nu} 
I_{log}(\mu^2) - 4 \Theta_{\mu \nu}^{(0)}$, etc.  
Then it follows that  
\begin{eqnarray}    
&&\Pi^{ab}_{\mu \nu}(1)+\Pi^{ab}_{\mu \nu}(2)+\Pi^{ab}_{\mu \nu}(3)   =  \nonumber \\
&& = g^2 C_2(G) \delta^{ab} (p^2g_{\mu \nu} -p_\mu p_\nu) \nonumber \\ && \times 
\Big[- b \frac 29 + \frac 53 \Big(I_{log}(\mu^2)- b Z_0 \Big) \Big]  .  
\end{eqnarray}     
The fermion loop contribution to the gluon self energy is identical to the vacuum 
 polarization tensor of $QED$ except for the colour and number of fermions ($N_f$) 
factors. It has been computed within IR elsewhere \cite{PRD2}. Here we only write 
the result in which we have already subtracted  
$I_{log}(\lambda^2)$ (that is to say we have employed the minimal subtraction in IR) 
and set the consistent relations to zero (CIR):  
\begin{eqnarray} 
&& \Pi_{\mu \nu}^{ab}(4)=-\frac i{144 \pi^2}g^2 \frac{N_f}{2}
\delta^{ab}(p_\mu p_\nu-p^2 g_{\mu \nu}) \times \nonumber \\ 
&& \times \Big( 12 Z_0 (m_f^2,m_f^2,p^2;\lambda^2) +4 \Big) \,\, .  
\end{eqnarray}  
Now the limit where $\mu \rightarrow 0$ can be safely taken because 
using $(\ref{eq:scarel})$ and that   
\begin{equation} 
Z_0(\mu^2,\mu^2,p^2;\mu^2) = \ln \frac{\lambda^2}{\mu^2} 
+Z_0(\mu^2,\mu^2,p^2;\lambda^2)\, ,  
\end{equation}  
one can show that  
\begin{eqnarray} 
&& I_{log}(\mu^2) - b Z_0(\mu^2,\mu^2,p^2;\mu^2) = I_{log}(\lambda^2) - \nonumber \\
&& - b Z_0(\mu^2,\mu^2,p^2;\lambda^2) \, ,  
\end{eqnarray}  
and  
\begin{equation} 
 Z_0(0,0,p^2;\lambda^2) = \ln \frac{p^2}{\lambda^2} - 2 \, .  
\end{equation}  
Let us also take the limit of massless fermions. Thus we have,   
\begin{equation}  
\Pi_{\mu \nu}^{ab}(4) =  
\frac i{144 \pi^2}g^2 \frac{N_f}{2}\delta^{ab}(p_\mu p_\nu-p^2 
g_{\mu \nu}) [12 \mbox{ln}(\bar {\lambda}^2/p^2)                
-4]  \, .  
\end{equation}  
Bringing  all the results together enables us to write the complete 
gluon self energy to one loop order, after minimally subtracting in the sense 
of IR the divergence expressed by $I_{log}(\lambda^2)$ and setting 
$\Upsilon_{\mu \nu}^0$ to zero as  
\begin{eqnarray}    
&& \Pi_{\mu \nu}^{ab}=\frac i{144 \pi^2}g^2 \delta^{ab}(p_\mu p_\nu-p^2 
g_{\mu \nu}) \times  \nonumber \\ &&\Big((15 C_2(R)- 6 N_f)\mbox{ln}
(\bar {\lambda}^2/p^2) -2C_2(R)+ 2N_f\Big)    
\end{eqnarray}    
which is just the result obtained in DfR \cite{PV3} after identifying 
$\lambda$ with the differential renormalization arbitrary scale ${\bar{M}}$ . Notice that 
both the massive and massless cases can be straightforwardly handled in IR.  
  
\section{Conclusions}  
  
In this paper we compared three frameworks namely differential (DfR),
 implicit (IR) and BPHZ regularisation/renormalisation which  being
 strictly defined in the physical dimension of the underlying theory
 may overcome the problems which arise when applying dimensional
 regularisation and variants to dimension specific theories such as
 supersymmetric, topological or chiral quantum field theories.

The  purpose was to motivate the answer to a few questions related to
the consistency and applicability of IR in handling
infinities in Feynman diagram calculations, namely: 1)
study how infrared divergences are treated within this scheme; 2)
understand how (nonabelian) gauge invariance can be automatically
implemented within a constrained version of IR; 3) define what a minimal
subtraction renormalisation is in IR in analogy with dimensional and
differential renormalization; 4) argue on the equivalence between IR, DfR and
DRed to one loop order and 4) motivate IR as a practical and
consistent tool for loop calculations in dimension specific theories.

Since in implicit regularization the divergences are displayed in the
form of basic divergent integrals it is natural to ask what is meant
by minimal subtraction in such scheme. We have shown that the
logarithmic divergences expressed by $I_{log}(\mu^2)$ can be split as
in equation $(\ref{eq:scarel})$ to give rise to am arbitrary scale
which plays the role of renormalisation scale in the Callan-Symanzik
equation. By subtracting $I_{log}(\lambda^2)$ when a logarithmic
divergence occurs we have a finite part which is identical to the
result in differential renormalization (with $\lambda$ playing the
role of the arbitrary scale $M$ in DfR) and dimensional regularisation
(except for a local counterterm). We define such renormalisation
scheme in IR as minimal. In constrained the arbitrary scale is also
taken from the logarithmic divergences only \cite{PV3C}(functions with
singular behaviour worse than logarithmic ($x^{-4}$) are reduced to
derivatives of less singular functions without introducing extra
dimensionful constants. This is the main ingredient that fixes the
renormalization scheme  in (constrained) DfR and automatically
preserves gauge invariance at least to one loop order.  We showed in
the calculation of the gluon selfenergy that a constrained
version of IR preserves gauge invariance just as it does for abelian
gauge theories \cite{PRD1}. Constraining IR amounts to set 
some well defined finite
differences between divergent integrals of the same degree of
divergence to zero \cite{PRD1}, \cite{PRD2}.   Such differences are
called consistency relations and were shown to be connected to
momentum routing invariance in the loops.
  
In order to illustrate the relationship between the BPHZ and the IR
schemes we have computed the setting sun diagram. The BPHZ scheme is a
very powerful tool for all order proofs, for instance, and it delivers
unambiguously the terms which ought to be subtracted in order to
render the amplitude finite by means of the forest formula. It is a
subtraction method which is regularisation independent in what
concerns the finite part. In order to proceed the calculation so to
extract the finite part, one has to apply ultimately a regularization
scheme. However symmetries may be broken in the course of such
operations such as gauge invariance. As we have seen within IR certain
surface terms are important in order to preserve gauge
invariance. Moreover, an expansion around zero in the external
momentum potentially breaks the gauge invariant structure of the
underlying amplitude. By controlling surface terms and using an
identity at the level of the integrand (\ref{eq:rr}) we have verified
that IR has control upon gauge invariance at least to one loop
order. In other words
in IR, the finite part is delivered automatically and no damage is
made to the integrand whereas  arbitrary local terms are duly parametrised in
IR by the consistency relations. A proof of renormalisability to all
orders in an alternative fashion to BPHZ has been constructed for
$\varphi^3_6$ theory within IR.

Although we have restricted ourselves to the gluon selfenergy, surely
one should verify if the other Slavnov-Taylor identites are satisfied
as well, by calculating the vertex functions of $QCD$ \cite{WIP}. 

Finally we conclude that both DfR and IR are potentially good
frameworks to apply in dimension specific problems in order o avoid
ambiguities and spurious anomalies. IR is particularly friendly from
the calculational viewpoint with the advantage of working directly in
the momentum space.  We believe that computations beyond one loop
order in Chern-Simons matter theories as well as in supersymmetric
models may profit from IR since such method does not modify the
underlying theory and operates in the physical dimension of the the
theory.

\section*{Appendix}   
The following relations  between the functions $Z_k$ and 
$\xi^{mn}$ can be easily checked and    
simplify the explicit verification of the Ward identity involving 
the electron self energy and the vertex function.   
   
\begin{equation}   
q^2 \xi_{01} - p.q \xi_{10} = {1\over 2} \{Z_0 (q^2;   
m^2) -   
Z_0 (p.q; m^2) + p^2 \xi_{00}\}   
\end{equation}   
\begin{eqnarray}   
q^2\xi_{11} - p.q\xi_{20} &=& {1\over 2} \Bigg\{{-Z_0(p +   
q)^2;m^2)   
\over 2}   
+ {Z_0(p^2; m^2)\over 2} \nonumber \\
&+& q^2 \xi_{10}\Bigg\}   
\end{eqnarray}   
\begin{equation}   
q^2\xi_{02} - p.q\xi_{11} = {1\over   
2}\left\{-\left[{1\over 2} + m^2   
\xi_{00}\right] + {p^2\over 2} \xi_{10} + {3q^2\over 2}   
\xi_{01}\right\}   
\end{equation}   
\begin{equation}   
p^2\xi_{20} - p.q\xi_{11} = {1\over 2}   
\left\{-\left[{1\over 2}   
+ m^2 \xi_{00}\right] + {q^2\over 2} \xi_{01} +   
{3p^2\over 2} \xi_{10}\right\}   
\end{equation}   
\begin{eqnarray}   
&& p^2\xi_{11} - p.q\xi_{02} = {1\over 2} \Big(-{1\over 2}   
Z_0((p+q)^2;m^2) \nonumber \\ && + {1\over 2} Z_0(q^2;m^2) + p^2   
\xi_{01}\Big) \, .   
\end{eqnarray}   
\begin{eqnarray}     
&& \widetilde{Z}(\mu_\gamma^2,m^2,p,q) = 1/2 Z_0((p-q)^2,m^2)-(1/2+\mu_\gamma^2 \xi^{00})   
\nonumber \\ && + 1/2 (q^2+\mu_\gamma^2-m^2)\xi^{10} + 1/2 (p^2+\mu_\gamma^2-m^2)\xi^{01} \, .   
\end{eqnarray}   
where $\xi^{mn} = \xi^{mn}(m^2,m^2,p,q) $ and we have abbreviated   
\begin{equation} 
Z_k (m^2,m^2,p^2;m^2) \equiv Z_k (p^2;m^2).    
\label{eq:abr}  
\end{equation}

\section*{Acknowledgements}   
The authors thank  Marcelo Gomes and  \'Elcio Abdalla for enlightening
 discussions. MS is grateful to the Mathematical Physics Department of S\~ao Paulo 
University for their hospitality. This work is supported by FCT/MCT/Portugal under
 the grant numbers PRAXIS XXI-BPD/22016/99, PRAXIS/C/FIS/12247/98, POCTI/1999/FIS/35309,
CNPq and Fape\-mig/Brazil.


\end{document}